# Gallium Nitride (GaN) based High-Power Multilevel H-Bridge Inverter for Wireless Power Transfer of Electric Vehicles

Javad Chevinly[1], Shervin Salehi Rad[1], Elias Nadi[2], Bogdan Proca[3], John Wolgemuth[3], Anthony Calabro[3], Hua Zhang[2], Fei Lu[1]

[1]Department of Electrical and Computer Engineerng, Drexel University, Philadelphia, PA, USA

[2]Department of Electrical and Computer Engineerng, Rowan University, Glassboro, NJ, USA

[3]InductEV Wireless Charging, King of Prussia, PA 19406

fei.lu@drexel.edu

***Abstract*—This paper presents a design and implementation of a high-power Gallium Nitride (GaN)-based multilevel H-bridge inverter to excite wireless charging coils for the wireless power transfer of electric vehicles (EVs). Compared to the traditional conductive charging, wireless charging technology offers a safer and more convenient way to charge EVs. Due to the increasing demand of fast charging, high-power inverters play a crucial role in exciting the wireless charging coils within a wireless power transfer system.**

**This paper details the system specifications for the wireless charging of EVs, providing theoretical analysis and a control strategy for the modular design of a 75-kW 3-level and 4-level H-bridge inverter. The goal is to deliver a low-distortion excitation voltage to the wireless charging coils. LTspice simulation results, including output voltage, Fast Fourier Transform (FFT) analysis for both 3-level and 4-level H-bridge inverters, are presented to validate the control strategy and demonstrate the elimination of output harmonic components in the modular design. A GaN-based inverter prototype was employed to deliver a 85-kHz power to the wireless charging pads of the wireless power transfer system. Experimental results at two different voltage and power levels, 100V-215W and 150V-489W, validate the successful performance of the GaN inverter in the wireless charging system.**

## I. Introduction

Wireless charging is becoming a popular option to achieve safe and convenient charging of electric vehicles (EVs). As the world embraces cleaner and more renewable energy, this technology is simplifying the charging process for EVs, offering a more user-friendly and sustainable transportation solution [1]-[11]. Modern wireless charging utilizes high frequency magnetic fields to transfer power resonantly, which is also the focused technology in this paper.

The power inverter (a power electronics converter) stands as a crucial component in EV wireless charging systems, especially for the fast-charging applications. Given the high-power requirements of EVs to realize rapid charging, the wireless charging system and wireless pads rely on a robust power inverter to receive the necessary high-power and high -frequency excitation from the transmitter side, making the presence of a high-power inverter indispensable for efficient wireless charging [6]-[9].

In practical application, the input voltage needs to be increased to achieve high power charging. Usually, the medium voltage power grid is directly used to supply power to a fast-charging station. For example, the input dc voltage can be a few kV. In this paper, the dc voltage is selected as 1.5 kV as a case study. Considering the high voltage levels at the input side, which stem from the substantial power transfer required for wireless charging, the stress on each inverter switch becomes a critical consideration [3]-[5]. There are several solutions available to address these challenges, with multilevel inverters being one of the practical options for high power, particularly when considering load sharing between each layer and other key system specifications [10].

To address the challenges, the paper introduces a solution by employing a multilevel inverter and incorporating multiple H-bridge inverters in a series connection at the transmitter side. This configuration is designed to reduce the stress voltage on each switch, offering a practical way to mitigate potential safety issues of high-power operation. The use of a multilevel inverter presents additional benefits, including the elimination of high-order harmonic components.

Specifically, this paper introduces a high-power multilevel inverter structure using gallium nitride (GaN) devices for wireless power transfer (WPT) systems in EVs. The system specification of a high-power WPT system is described in detail with three-level and four-level H-bridge configurations to ensure a low-distortion output voltage and current for wireless charging pads. Validation of the control strategy and output voltage is demonstrated through LTSpice simulations for a 75-kW high-power inverter example. Additionally, this paper presents the experimental results of GaN inverters in wireless charging for different power level to validate the inverter performance. Further research will implement high power GaN inverters to achieve fast charging for EV applications.

## II. SYSTEM SPECIFICATION OF WPT FOR EVS

In this section, this paper illustrates the system-level structure of a wireless charger for EVs and outlines the system specifications for the GaN-based multilevel H-bridge inverter designed for the wireless charging system.

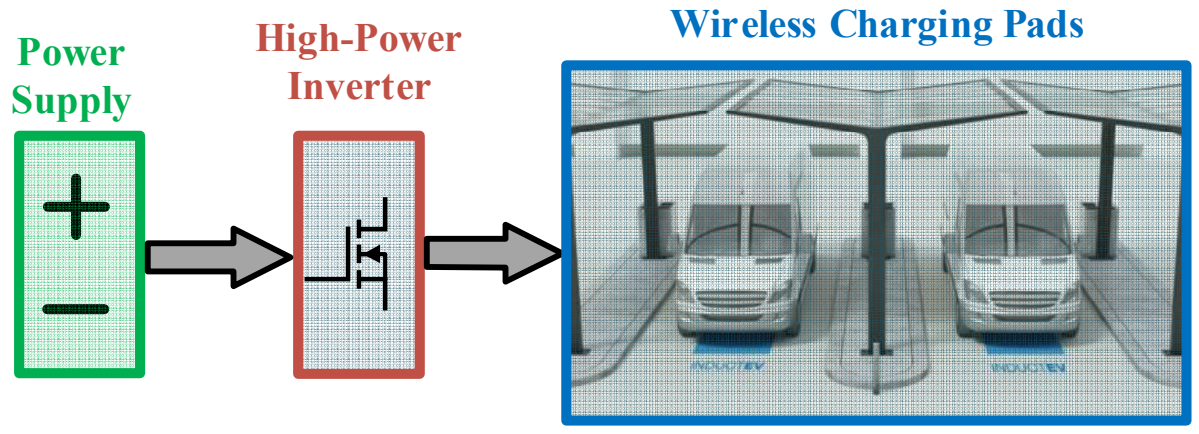

(a) Wireless charging system of EV.

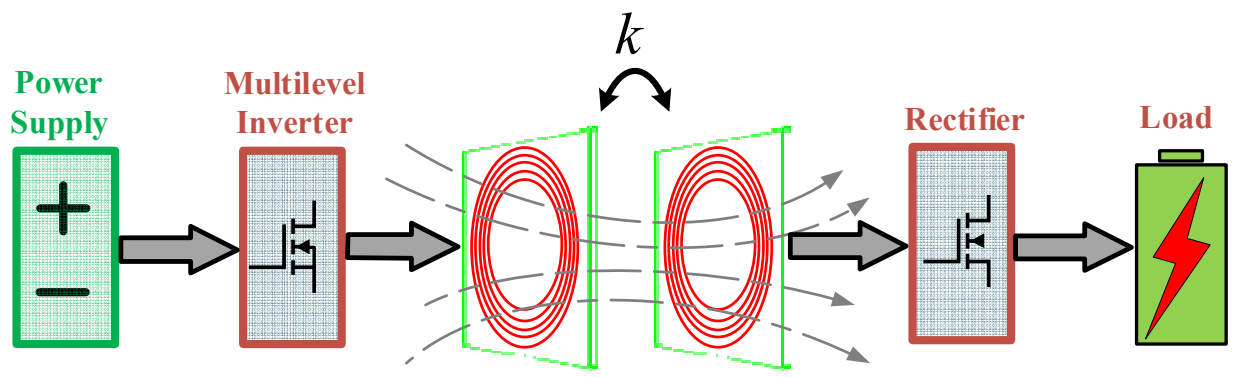

(b) Wireless power transfer system

**Figure 1.** Concept structure of the system configuration of the proposed high-power wireless power transfer for EVs.

Figure 1(a) depicts the high-power inverter connected to wireless charging pads for EVs, emphasizing the critical role of the inverter in facilitating efficient wireless charging. Figure 1(b) details the wireless power transfer for EVs, presenting both the transmitter and receiver sides of the system. The high-power multilevel inverter generates the excitation voltage necessary to achieve wireless charging. A rectifier is required to charge the battery on the receiver side of the EV. This dual-component system ensures an effective wireless power transfer process for EVs with high efficiency. In this system structure, the high-power inverter technology is the focus of this paper's design and implementation target.

The block-diagram of a wireless charging system using a GaN based multilevel inverter is shown in Figure 2. The targeted total harmonic distortion (THD) for the output waveform is set to be less than 10%. Adhering to the SAE-J2954 standard (79~90 kHz), the operational switching frequency of the multilevel inverter is optimized within the 85-kHz range to minimize switching losses. Moreover, the GaN-based multilevel H-bridge inverter is crafted to operate with an efficiency exceeding 98.5%, ensuring an effective and energy-efficient performance. The efficiency consideration is mainly based on the thermal performance of the GaN device and the heatsink design for the inverter. Based on the relatively small package of GaN devices, the efficiency can be higher to maintain a safe operation.

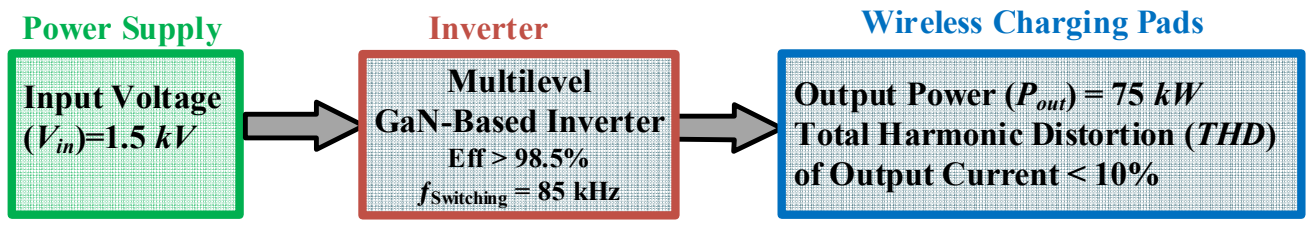

**Figure 2.** Block diagram of the proposed wireless charging system (ground side structure).

## III. THEORETICAL ANALYSIS AND CONTROL STRATEGY OF MULTILEVEL H-BRIDGE INVERTER

In this section, a theoretical analysis of a multilevel H-bridge inverter is performed. The control strategy is aimed at canceling harmonic components while achieving load sharing between the layers of the multilevel inverter. The analysis is done for both three-level and four-level inverters.

### A. *Three-level H-bridge inverter*

In a typical multi-level inverter design, the input voltage is distributed among multiple H-bridge inverters in a series connection. For example, the topology structure of a 3-level inverter is provided in Figure 3.

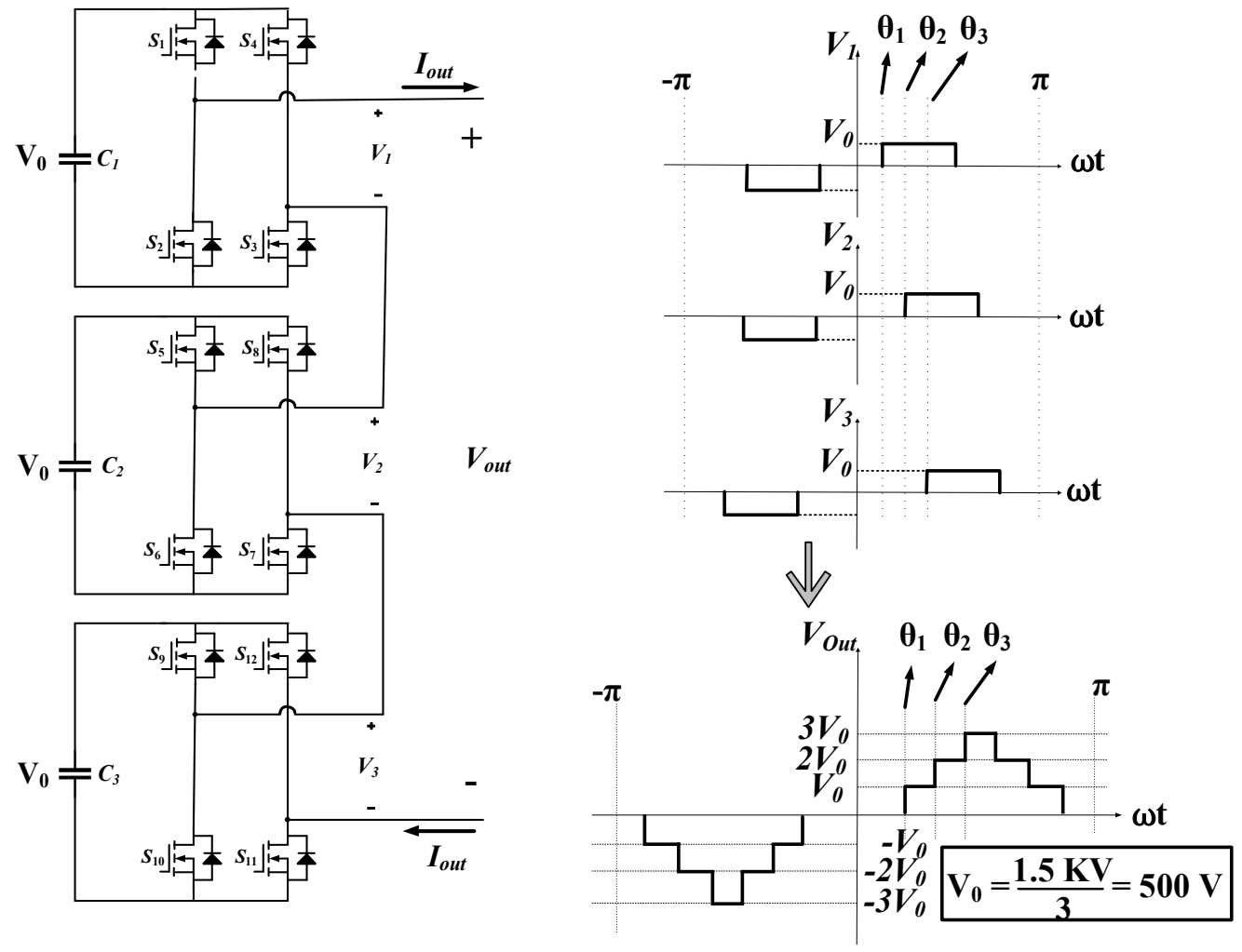

**Figure 3.** Proposed three-level GaN-based inverter topology to reduce harmonics in the ac excitation voltage.

As shown in Figure 3, $V_{DC}$ represents the total input voltage. In this case, it is set at 1.5 kV. The output voltages of each H-bridge inverter layer, denoted as $V_1$, $V_2$, and $V_3$, respectively, exhibit square waveforms as anticipated. In this example, the phase angles $\theta_1$, $\theta_2$, and $\theta_3$ are associated with each H-bridge layer, individually defined by the switches of each inverter. The overall output voltage of the multilevel inverter ($V_{out}$) is the sum of the outputs of each inverter layer ($V_1$, $V_2$, and $V_3$), aligned in a series connection. The design of these inverter layer angles, represented by $\theta_1$, $\theta_2$, and $\theta_3$, is critical for achieving load sharing and canceling harmonic components. Applying FFT to $V_{out}$ of the multilevel inverter is also an important step in the analysis.

To eliminate the 3rd, 5th, and 7th harmonic components in this three-level inverter, equations (1), (2), and (3) below need to be satisfied.

$$\cos(3\theta_1) + \cos(3\theta_2) + \cos(3\theta_3) = 0 \quad (1)$$
$$\cos(5\theta_1) + \cos(5\theta_2) + \cos(5\theta_3) = 0 \quad (2)$$
$$\cos(7\theta_1) + \cos(7\theta_2) + \cos(7\theta_3) = 0 \quad (3)$$
$$\Rightarrow \quad \theta_1 = 11^\circ, \quad \theta_2 = 41^\circ, \quad \theta_3 = 85^\circ$$

Solving these three nonlinear equations defines the values of the three angles, $\theta_1$, $\theta_2$, and $\theta_3$, ensuring effective load sharing and harmonic component cancellation in the system.

### B. *Four-level H-bridge inverter*

In this section, the theoretical analysis and control strategy for a four-level inverter are discussed. The input voltage is distributed among four H-bridge inverters in a series connection. The topology structure of the 4-level inverter is illustrated in Figure 4.

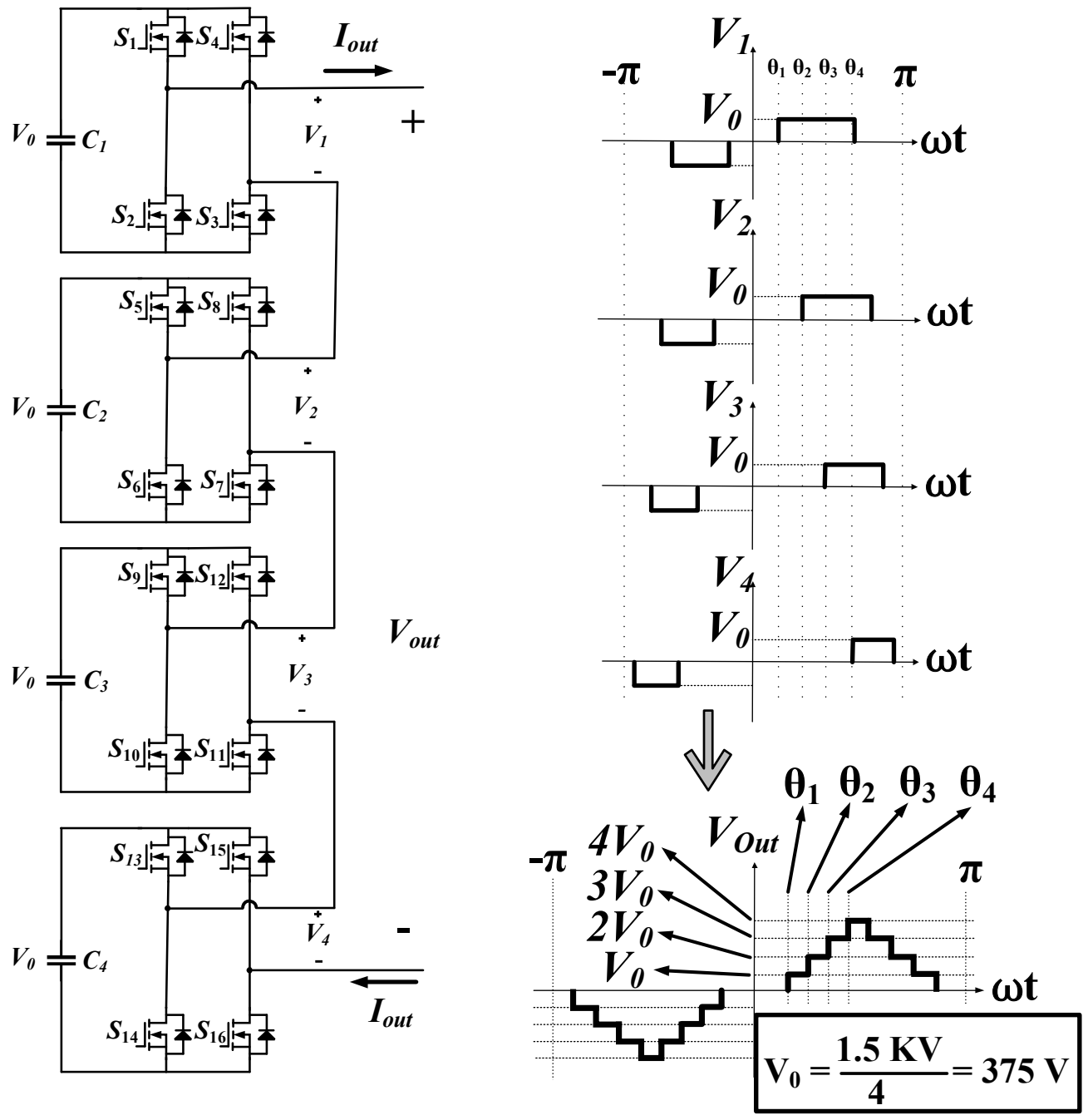

**Figure 4.** Proposed four-level GaN-based inverter topology to reduce harmonics in the ac excitation voltage.

In Figure 4, $V_{DC}$ is the input voltage, which is consistent with the previous case. The output voltages of each H-bridge inverter layer, denoted as $V_1$, $V_2$, $V_3$, and $V_4$, exhibit square waveforms. Angles $\theta_1$, $\theta_2$, $\theta_3$, and $\theta_4$ are associated with each layer, defined by the respective inverter layers. The overall output voltage of the four-level inverter ($V_{out}$) is the sum of the outputs of each inverter layer in a series connection.

To eliminate the 3rd, 5th, 7th, and 9th harmonic, equations (4), (5), (6), and (7) are employed. $\theta_1$, $\theta_2$, $\theta_3$, and $\theta_4$ are determined by solving nonlinear equations.

$$\cos(3\theta_1)+\cos(3\theta_2)+\cos(3\theta_3)+\cos(3\theta_4)=0 \quad (4)$$
$$\cos(5\theta_1)+\cos(5\theta_2)+\cos(5\theta_3)+\cos(5\theta_4)=0 \quad (5)$$
$$\cos(7\theta_1)+\cos(7\theta_2)+\cos(7\theta_3)+\cos(7\theta_4)=0 \quad (6)$$
$$\cos(9\theta_1)+\cos(9\theta_2)+\cos(9\theta_3)+\cos(9\theta_4)=0 \quad (7)$$

$$\Rightarrow \quad \theta_1=9^\circ,\ \theta_2=26^\circ,\ \theta_3=50^\circ,\ \theta_4=86^\circ$$

## IV. CIRCUIT SIMULATION RESULTS BY LTSPICE

### A. *LTspice simulation result of three-level H-bridge inverter*

Simulations are conducted for the multilevel H-bridge inverter operating with an input voltage of 1.5 kV and a frequency of 85 kHz using LTspice. The simulation results for the 3-level inverter are presented in Figure 5.

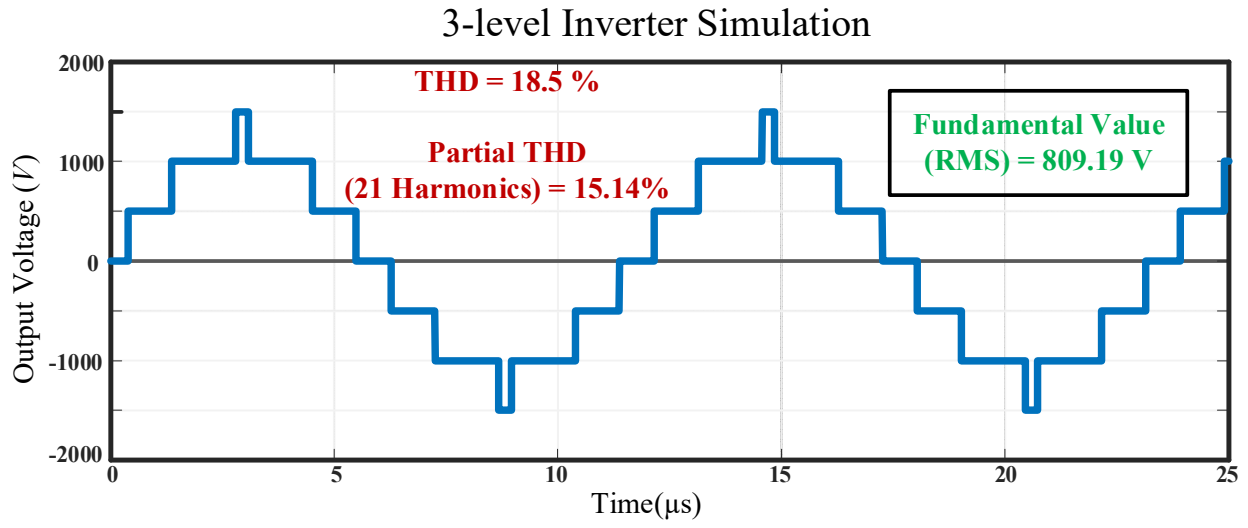

(a) Output Voltage waveform.

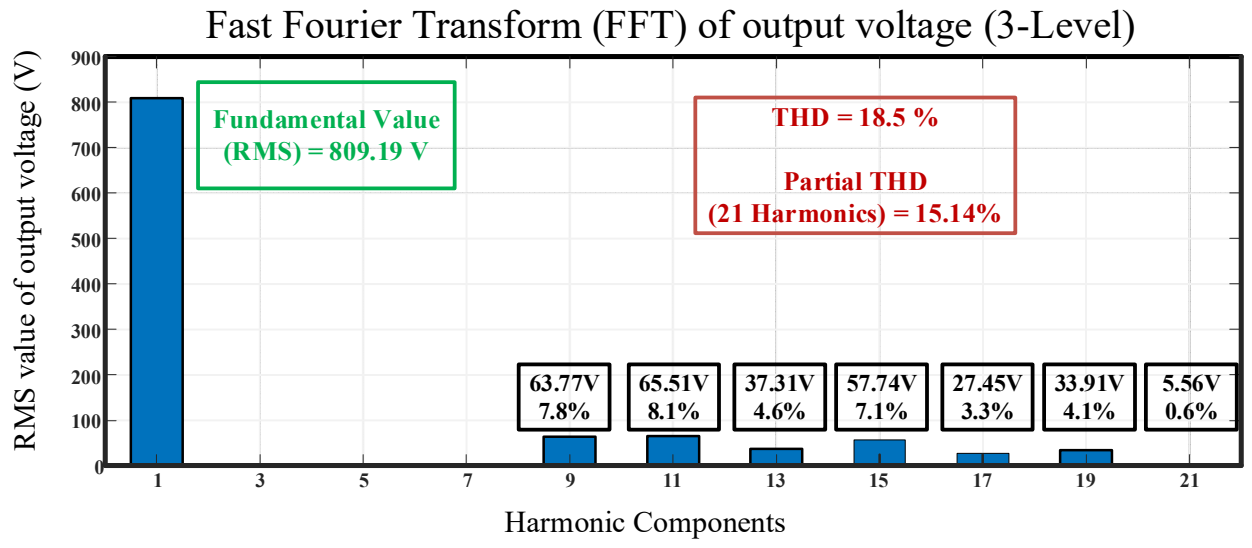

(b) FFT plot of output voltage

**Figure 5.** LTSpice simulation result of the proposed three-level H-bridge inverter with the calculated phase angle $\theta_1$, $\theta_2$, and $\theta_3$.

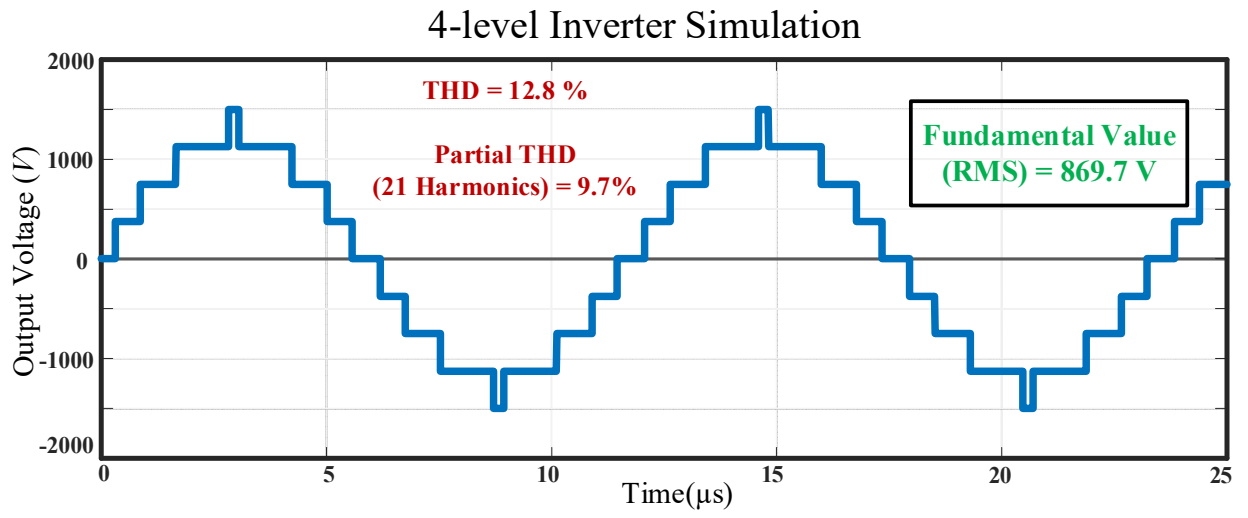

(a) Output Voltage waveform.

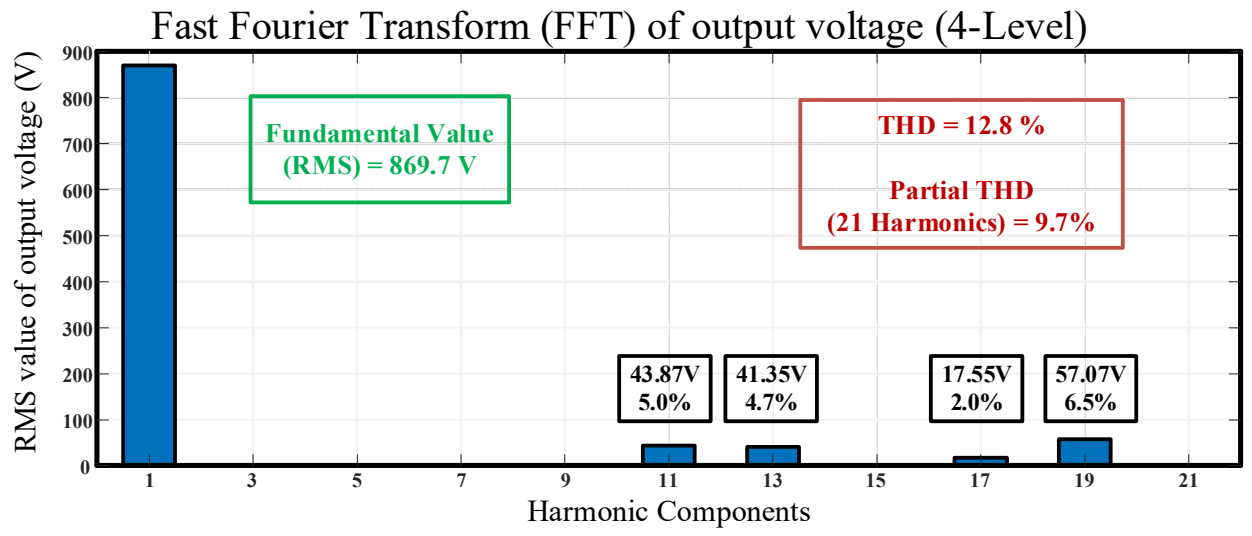

(b) FFT plot of output voltage

**Figure 6.** LTspice simulation result of the proposed four-level H-bridge inverter with the calculated phase angle $\theta_1$, $\theta_2$, $\theta_3$, and $\theta_4$.

In Figure 5, the calculated phase angles for each inverter layer are provided. Figure 5(a) illustrates the output voltage waveform for the 3-level inverter, where the input voltage for each H-bridge inverter is set at 500V. The RMS value of the fundamental output voltage is measured at 809.19V.

In Figure 5(b), the FFT plot of the output voltage demonstrates the successful elimination of the 3rd, 5th, and 7th harmonic components, aligning with the technical design objectives. The THD for the output voltage waveform is approximately 18.5%, and for the first 21 harmonics, the THD is around 15.14%. This validates that the calculated phase angles are effective at canceling harmonics, and that the output waveform has low high-order harmonics..

## B. *LTspice simulation result of four-level H-bridge inverter*

In Figure 6(a), the simulation illustrates the output voltage of a four-level inverter. The input voltage for each is 375V, as it is evenly divided among four inverters. The entire RMS value of the output voltage is 869.7V.

Figure 6(b) presents the results of FFT plot, validating the successful elimination of the 3rd, 5th, 7th, and 9th harmonics. The THD for the output voltage waveform is approximately 12.8%, and for the first 21 harmonics, the THD is around 9.7%. Compared to the previous three-level inverter in Figure 5, this shows that the total harmonics content is significantly reduced. Since the voltage stress on each inverter is also reduced, this proves that increasing the number of voltage levels helps to provide a higher performance inverter to excite WPT.

# V. Experimental Validation and Results

## A. *Implementation of a GaN-based H-Bridge Inverter*

An H-bridge inverter is implemented based on GaN devices to experimentally validate the proposed inverter structure and function. The hardware prototype and details of an H-bridge inverter are shown in Figure 7.

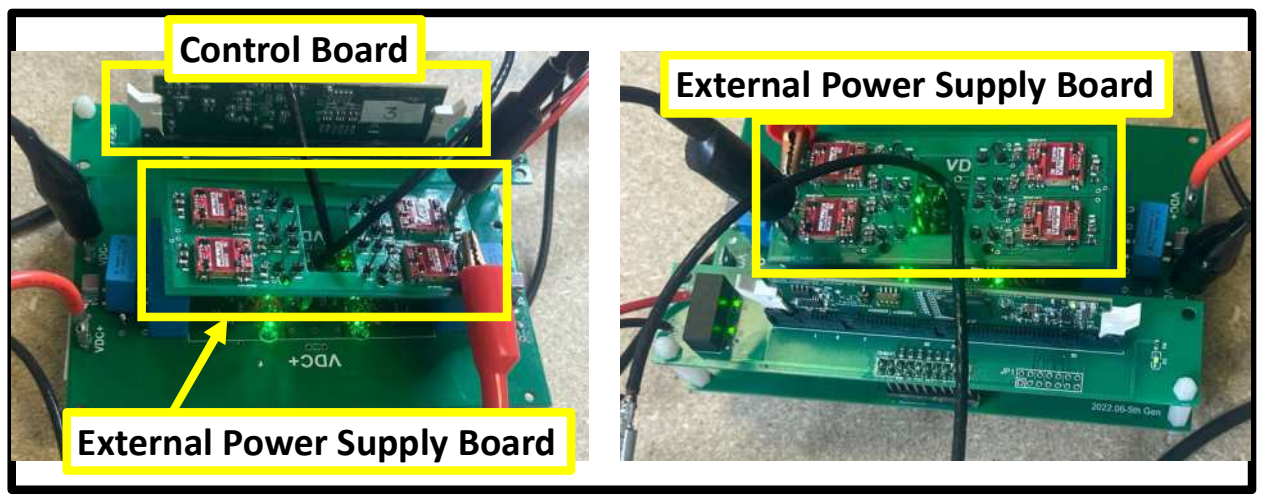

**Figure 7.** Hardware implementation of GaN-based Inverter.

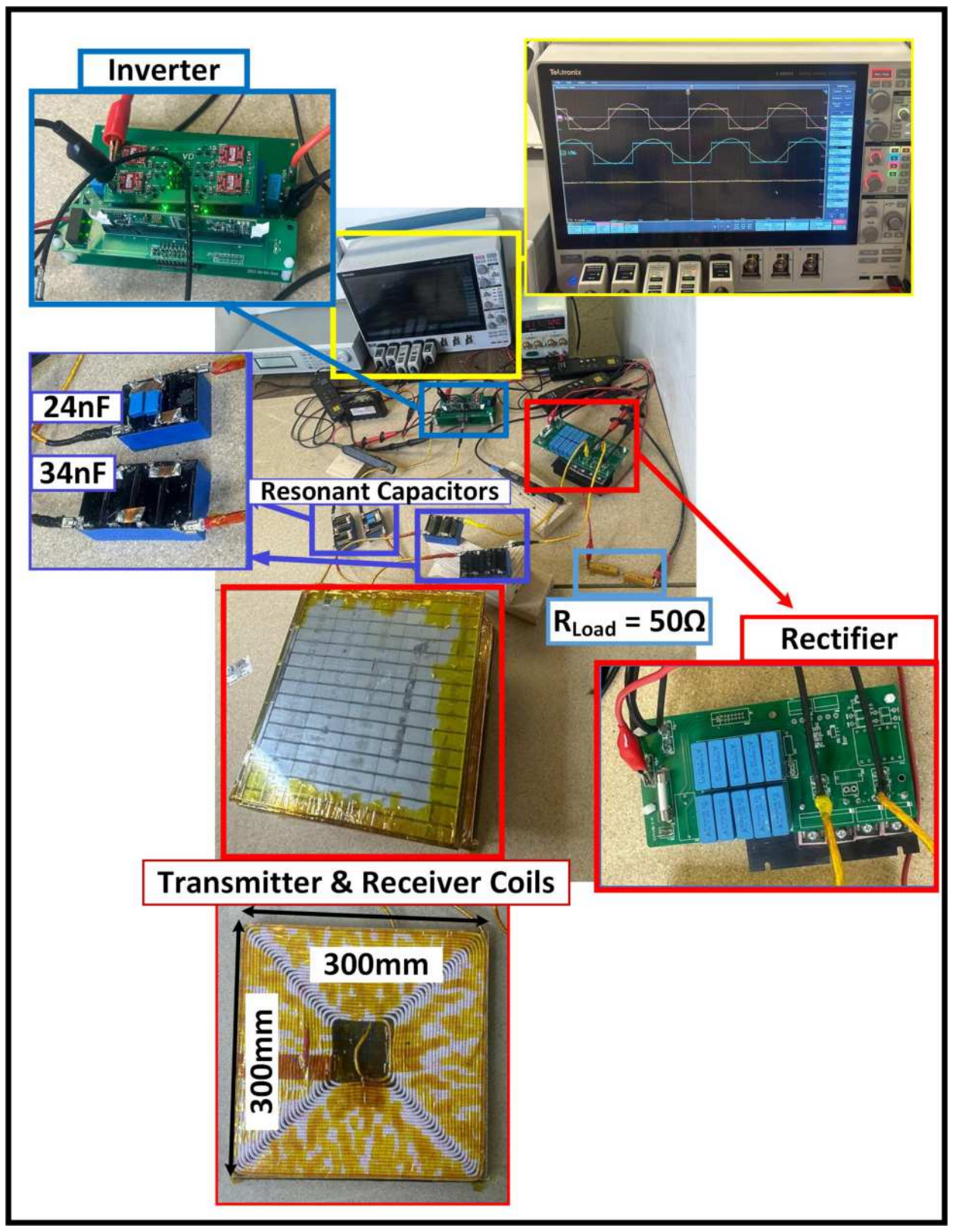

**Figure 8.** Experimental setup of the GaN-based inverter in a WPT system.

**Table I:** The implemented WPT System specifications.

| Parameter | Value | Parameter | Value |
|---|---|---|---|
| Input $V_{dc}$ | 100V-150V | resonant frequency $f_l$ | 85kHz |
| GaN Device | GS66506T | Airgap | 105mm |
| Gate Driver | UCC5350SB | $L_1 = L_2$ | 245uH |
| DSP | MS320F28335 | $C_1 = C_2$ | 14nf |
| $k$ (well aligned) | 0.309 | Load | 50Ω |
| Switching frequency $f_s$ | 85kHz | | |

In Figure 7, there are three main sections highlighted: the control board, which utilizes the DSP TMS320F28335 to generate pulses for controlling the GaN switches; the external power supply board, which employs power modules to supply the gate drivers; and the main PCB board, which contains input capacitor filters, switch drivers, anti-parallel diodes, and GaN devices. Additionally, a heatsink has been designed to address the thermal concerns of the GaN devices. This heatsink is installed on the bottom side of the circuit board with an insulation layer between the GaN devices. Specifically, GS66506T GaN transistors ($V_{DS}$ = 650 V, $R_{DS(on)}$ = 67 mΩ, $I_{DS(max)}$ = 22.5 A) are utilized for this multilevel inverter. Furthermore, the anti-parallel diode E4D02120E-TR is incorporated into this GaN inverter implementation.

## B. *Experimental Testing of a GaN-based H-Bridge Inverter*

Using the implemented H-bridge GaN inverter, further experiments are conducted to test its power conversion performance. During the test process, Figure 8 further illustrates the setup of a series-series compensated inductive power transfer system for this GaN-based inverter. The GaN inverter is connected to the transmitter coil via primary resonant capacitors, and the receiver coil is connected to a rectifier via secondary resonant capacitors. A Schottky diode voltage rectifier is constructed on the secondary side. The output of the rectifier on the receiver side is connected to the DC load, which is 50Ω.

Table I presents the parameters of the implemented wireless power transfer system. The system was tested with input DC voltages ranging from 100 V to 150 V across two voltage levels, with the switching frequency ($f_s$) set to 85 kHz (equal to the resonant frequency ($f_l$)). The primary and receiver coils share identical inductance values of 245 μH, and the primary and secondary resonant capacitors are both set at 14 nF. In the well-aligned situation, with an air gap of 105 mm, the coupling coefficient ($k$) is measured at 0.309. The performance of the wireless power transfer system with the GaN inverter was evaluated through experimental testing.

Figure 9 provides the experimental results of the GaN inverter within the proposed WPT system for supplying the load at two voltage and power levels. At a 100 V DC input voltage, it delivers 215 W at the output side, while at a 150 V DC input voltage, it supplies 489 W to the load side, both under well-aligned conditions with a coupling coefficient of 0.309. Additionally, the zero-voltage switching (ZVS) operation is achieved in both operating conditions due to resonant operation. The experiments have demonstrated that the implemented GaN inverter can provide high-voltage excitation to achieve wireless power transfer. In future work, multiple H-bridge inverters will be further implemented and connected in series to achieve high power capability up to tens of kW for fast charging applications.

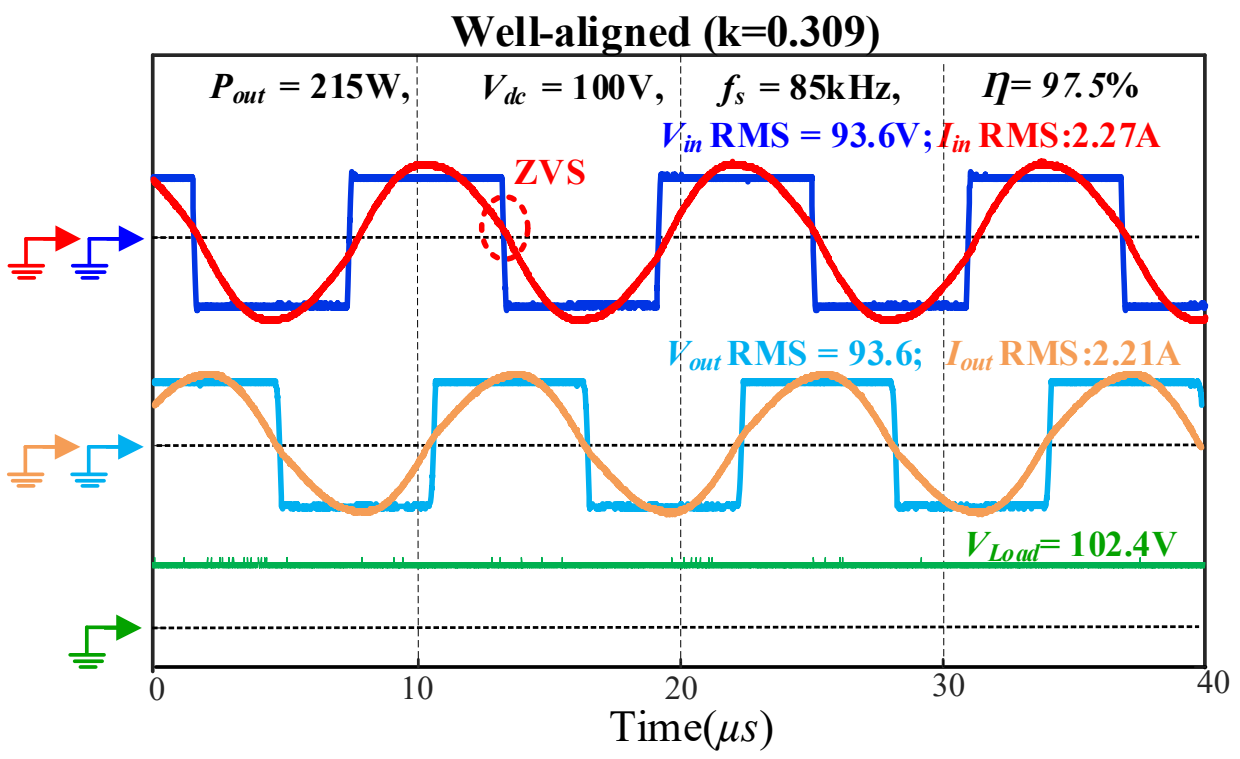

(b)

**Fig .9.** Experimental waveforms well-aligned *k*=0.309 (a) $V_{dc}$=100V, P=215W (b) $V_{dc}$=150V, P=489W

## VI. Conclusions

This paper presents a high-power GaN-based multilevel H-bridge inverter specifically designed for wireless charging applications. Designs for both 3-level and 4-level H-bridge inverters have been explored to achieve the high output voltages necessary to power wireless charging systems at high power levels. The control strategy includes mechanisms for load sharing and methods for eliminating harmonics, focusing on the 3$^{rd}$, 5$^{th}$, and 7$^{th}$ orders for the 3-level inverter, and the 3$^{rd}$, 5$^{th}$, 7$^{th}$, and 9$^{th}$ orders for the 4-level inverter. LTspice simulation results are presented to prove the methodology and demonstrate harmonic elimination for both the 3-level and 4-level inverters. A GaN inverter prototype is integrated into a WPT system to generate 85 kHz excitation for the primary coil and supply power to the load on the receiver side. Experimental results obtained at two different voltage and output levels (100V-215W and 150V-489W) validate the performance of the GaN inverter within the WPT system.